\documentclass[journal]{IEEEtran}
\usepackage{graphicx,color}

\usepackage[OT1]{fontenc} 
\usepackage[numbers,sort&compress]{natbib}
\usepackage[cmex10]{amsmath}
\usepackage{amssymb}
\usepackage{bm}
\usepackage{graphicx}
\usepackage{color}
\usepackage{tabularx}
\usepackage{arydshln}
\usepackage{mathtools}
\usepackage{multirow}
\usepackage{algorithm,algorithmic}
\usepackage{url}
\usepackage{listings}
\usepackage{comment}
\usepackage{CJKutf8}
\usepackage{subfigure}
\RequirePackage[normalem]{ulem}
\usepackage{hyperref}
\usepackage[absolute,overlay]{textpos}
\def\Xtf{\mathbf{X}_\mathrm{tf}}

\def\X{\mathbf{X}}
\def\x{\mathbf{x}}

\def\Ytf{\mathbf{Y}_\mathrm{tf}}

\def\Y{\mathbf{Y}}

\def\y{\mathbf{y}}

\def\grx{g_\mathrm{rx}}
\def\Gtx{\mathbf{G}_\mathrm{tx}}
\def\Grx{\mathbf{G}_\mathrm{rx}}
\def\FN{\mathbf{F}_\mathrm{N}}
\def\FM{\mathbf{F}_\mathrm{M}}

\def\IM{\mathbf{I}_\mathrm{M}}
\def\H{\mathbf{H}}
\def\Hhat{\hat{\mathbf{H}}}

\def\lmax{l_\mathrm{max}}
\def\kmax{k_\mathrm{max}}
\def\Np{N_p}

\def\O{\mathcal{O}}
\def\rend{r_\mathrm{end}}
\def\runc{r_\mathrm{unc.}}
\def\rcod{r_\mathrm{cod.}}

\makeatletter
\def\ps@IEEEtitlepagestyle{%
  \def\@oddfoot{\mycopyrightnotice}%
  \def\@oddhead{\hbox{}\@IEEEheaderstyle\leftmark\hfil\thepage}\relax
  \def\@evenhead{\@IEEEheaderstyle\thepage\hfil\leftmark\hbox{}}\relax
  \def\@evenfoot{}%
}
\def\mycopyrightnotice{%
  \begin{minipage}{\textwidth}
  \centering \scriptsize
\textcopyright 2025 IEEE. Personal use of this material is permitted.
  Permission from IEEE must be obtained for all other uses, in any current or future
  media, including reprinting/republishing this material for advertising or promotional
  purposes, creating new collective works, for resale or redistribution to servers or
  lists, or reuse of any copyrighted component of this work in other works.
  DOI: \href{https://doi.org/10.1109/TVT.2025.3608199}{10.1109/TVT.2025.3608199}
  \end{minipage}
}
\makeatother
\begin{document}
\title{Superimposed Pilot-Based OTFS: Will it Work?}

% Author
\author{
Yuta~Kanazawa,~Hiroki~Iimori,~\IEEEmembership{Member,~IEEE},
Chandan~Pradhan,~\IEEEmembership{Member,~IEEE},\\
Szabolcs~Malomsoky,
and Naoki~Ishikawa,~\IEEEmembership{Senior Member,~IEEE}.
\thanks{Y.~Kanazawa and N.~Ishikawa are with the Faculty of Engineering, Yokohama National University, 240-8501 Kanagawa, Japan (e-mail: ishikawa-naoki-fr@ynu.ac.jp). H.~Iimori, C.~Pradhan, and S.~Malomsoky are with Ericsson Research, Ericsson Japan K.K., 220-0012 Kanagawa, Japan.

A part of this paper was presented at the IEEE 98th Vehicular Technology Conference \cite{kanazawa2023multiple}.
}}

%\markboth{\today}
%{Shell \MakeLowercase{\textit{et al.}}: Bare Demo of IEEEtran.cls for Journals}
\maketitle

%ARXIV
\TPshowboxesfalse
\begin{textblock*}{\textwidth}(45pt,10pt)
\footnotesize
\centering
Accepted for publication in IEEE Transactions on Vehicular Technology. This is the author's version which has not been fully edited and content may change prior to final publication. Citation information: DOI 10.1109/TVT.2025.3608199
\end{textblock*}
%ARXIV

\begin{abstract}
Orthogonal time frequency space (OTFS) modulation is a promising solution to handle doubly-selective fading, but its channel estimation is a nontrivial task in terms of maximizing spectral efficiency. Conventional pilot assignment approaches face challenges: the standard embedded pilot-based scheme suffers from low transmission rates, and the single superimposed pilot (SP)-based scheme experiences inevitable data-pilot interference, leading to coarse channel estimation. To cope with this issue, focusing on the SP-based OTFS system in channel coded scenarios, we propose a novel pilot assignment scheme and an iterative algorithm. The proposed scheme allocates multiple SPs per frame to estimate channel coefficients accurately. Furthermore, the proposed algorithm performs refined interference cancellation, utilizing a replica of data symbols generated from soft-decision outputs provided by a decoder. Assuming fair and unified conditions, we evaluate each pilot assignment scheme in terms of reliability, channel estimation accuracy, effective throughput, and computational complexity. Our numerical simulations demonstrate that the multiple SP-based scheme, which balances the transmission rate and the interference cancellation performance, has the best throughput at the expense of slightly increased complexity. In addition, we confirm that the multiple SP-based scheme achieves further improved throughput due to the proposed interference cancellation algorithm.
\end{abstract}


\IEEEpeerreviewmaketitle

\section{Introduction \label{sec:intro}}

Orthogonal time frequency space (OTFS) modulation \cite{hadani2017orthogonal} and its related technique \cite{rou2024orthogonal,sui2023low,sui2023space} offer a promising solution for high-mobility communication scenarios, including high-speed trains and unmanned aerial vehicles.
OTFS modulates the information symbols on the delay-Doppler (DD) domain providing robust equalization for doubly-selective (time and frequency) fading caused by serious Doppler shifts.
However, OTFS modulation processes large information blocks across time and frequency resources, increasing latency and complexity \cite{aldababsa2024survey}.
Several detectors have been proposed for OTFS, such as message passing (MP) \cite{raviteja2018interference}, linear minimum mean square error \cite{tiwari2019low, surabhi2020low}, and maximal ratio combining \cite{thaj2020low}, each designed to exploit the sparsity and block circulant property of channel matrices, thereby offering low-complexity detection.
More recent approaches, such as variational Bayes \cite{yuan2020simple}, unitary approximate message passing \cite{yuan2022iterative}, and deep learning-based detection methods \cite{xu2020damped, enku2021two, naikoti2021low}, aim to enhance detection accuracy and computational efficiency.
While these studies assume perfect knowledge of channel state information (CSI), practical OTFS systems require effective channel estimation to account for real-world uncertainties.

Prior studies \cite{raviteja2019embedded, murali2018otfs, shen2019channel, qu2021low} assume channel estimation using known pilot symbols between the transmitter and the receiver, resulting in reduced throughput.\footnote{There existing another attractive approach that dispenses with channel estimation \cite{xu2024noncoherent}.}
In particular, the most typically considered embedded pilot (EP)-based scheme \cite{raviteja2019embedded} realizes low-complexity accurate channel estimation using a single pilot symbol surrounded by sufficient guard space \cite{wei2021orthogonal}.
In high-mobility scenarios, the EP-based scheme requires large guard space depending on a maximum Doppler shift to avoid data-pilot interference, which results in further decreased spectral efficiency.
An alternative approach to improve spectral efficiency is the superimposed pilot (SP) \cite{hoeher1999channel}, which overlays data and pilot symbols without requiring guard space, thereby maximizing transmission capacity.
Therefore the SP-based channel estimation scheme achieves high transmission rate at the cost of potentially lower channel estimation accuracy.
For example, a data-aided channel estimation algorithm is proposed in \cite{yuan2021dataaided}, which places a single SP symbol in the frame in the DD domain.
The single SP-based scheme realizes accurate channel estimation using an iterative detection process.

In order to improve reliability, channel estimation accuracy, and peak-to-average power ratio (PAPR), a number of SP-based OTFS channel estimation schemes \cite{yuan2021dataaided, mishra2022otfs, jesbin2023sparse, liu2023low, chen2023delaywise, liu2024turbo} have been proposed.
To introduce a few examples, in \cite{jesbin2023sparse}, Jesbin et al. proposed the SP-based scheme with sparsely arranged pilot symbols in the DD domain, which realized better reliability and channel estimation accuracy compared to the single SP-based scheme.
In \cite{liu2023low}, Liu et al. proposed a novel pilot assignment scheme with several SP symbols on the different delay axis in the DD domain, which realized PAPR reduction.
Furthermore, in \cite{chen2023delaywise}, Chen et al. proposed another SP-based scheme which arranged one pilot for each delay tap to reduce PAPR.
These studies \cite{yuan2021dataaided, jesbin2023sparse, liu2023low, chen2023delaywise} focused on evaluating each pilot assignment scheme in terms of bit error rate (BER) and normalized mean squared error (NMSE) of the estimated channel matrices, thus the advantage of the SP-based scheme in terms of spectral efficiency has not been clarified.
Exceptions can be found in \cite{mishra2022otfs,liu2024turbo}.
In \cite{mishra2022otfs}, Mishra et al. evaluated the SP-based scheme in terms of the spectral efficiency, but it assumed uncoded scenarios.
In \cite{liu2024turbo}, considering coded scenarios, Liu et al. proposed a sophisticated channel estimation algorithm with the Turbo equalization concept, which achieved better BER and spectral efficiency compared to benchmark schemes \cite{qu2021low, mishra2022otfs}, at the cost of computational complexity due to matrix operations for pilot cancellation and signal detection.
These prior studies \cite{mishra2022otfs, liu2024turbo} derived the spectral efficiency by signal-to-interference-and-noise ratio (SINR), assuming ideal continuous input signals which follows Gaussian distribution.

The fundamental limitation of SP-based schemes is the residual data-pilot interference in the high SNR region \cite{yuan2021dataaided, mishra2022otfs, liu2024turbo}.
This interference limits the achievable accuracy of channel estimation and system reliability, especially under strong channel conditions.
In addition, some SP-based schemes such as those in \cite{mishra2022otfs,jesbin2023sparse} require prior knowledge of the delay-Doppler shifts of the channel paths.
This requirement can be satisfied by adopting an EP-based scheme in the initial transmission frame, which reduces overall spectral efficiency.
To address this limitation, simultaneous estimation of the channel gain and the delay-Doppler shifts using threshold-based techniques, as suggested in \cite{raviteja2019embedded, yuan2021dataaided}, becomes essential.

Given this background, we propose an SP-based channel estimation scheme suitable for coded scenarios, which consists of a novel pilot assignment approach and an iterative algorithm for improved data-pilot interference cancellation and estimation of the delay-Doppler shifts.
Furthermore, we demonstrate the advantage of our proposed scheme, through numerical simulations in terms of effective throughput, which is a practical spectral efficiency metric for discrete input signals such as quadrature amplitude modulation (QAM).
The major contributions are summarized as follows.
\begin{enumerate}
\item 
We propose a novel frame structure with multiple superimposed pilots, which provides sufficient guard space among each pilot to avoid interference.
Averaging the estimated channel coefficients given by multiple pilot symbols, we can realize the channel estimation and interference cancellation simultaneously.
\item
Focusing on a property that the quality of interference cancellation depends on estimated data symbols, we propose an iterative algorithm for the SP-based OTFS, which utilizes the capability of the channel coding.
In our proposed algorithm, log likelihood ratios (LLRs) corresponding to the information bits are obtained from the channel decoder and used to generate soft-decision replicas of data symbols.
The error-corrected data symbols contribute to accurate channel estimation.
While our proposed algorithm requires additional operations, we can alleviate the increased complexity by switching two iterative processes: uncoded and coded iterations.
\item
Assuming a practical scenario with 5G compliant low density parity check (LDPC) codes, we show the advantage of the SP-based scheme in terms of BER, NMSE, and throughput.
Through the complexity analysis, we confirm that the impact on the increased computational complexity is negligible compared to the conventional scheme.
We conduct fair comparisons for the EP-based and SP-based schemes under a proper SNR definition which offers unified signal power.
\end{enumerate}

The rest of this paper is organized as follows.
The overview of the OTFS system is summarized in Section~\ref{sec:overview}, and the OTFS channel estimation schemes are reviewed in Section~\ref{sec:est}.
Section~\ref{sec:algo} introduces the proposed iterative algorithm for coded scenarios, including its computational complexity analysis.
The simulation results are shown in Section~\ref{sec:simu}.
Finally, we conclude this paper in Section~\ref{sec:conc}.

\textit{Notations}:
Throughout this paper, $(\cdot)^{*}$, $(\cdot)^\mathrm{T}$, and $(\cdot)^\mathrm{H}$ denote the complex conjugation, transposition, and Hermitian transposition, respectively. 
The notation $\mathrm{E}[\cdot]$ denotes the expectation operator.
The operation $\otimes$ represents the matrix Kronecker product and $\mathbf{a} = \mathrm{vec}(\mathbf{A})$ is the column-wise vectorization of a matrix $\mathbf{A}$ to a vector $\mathbf{a}$.
Contrary, $\mathbf{A} = \mathrm{vec}^{-1}(\mathbf{a})$ denotes the inverse vectorization of a vector $\mathbf{a}$ to form a matrix $\mathbf{A}$.
The notations $\mathrm{diag}[\cdot]$, $\delta (\cdot)$, $[\cdot]_n$, and $\|\cdot\|_\mathrm{F}$ express the diagonal matrix, the Dirac delta function, the modulo by an integer $n$, and the Frobenius norm of a matrix, respectively.
Finally, let $\mathbb{R, C}$, and $\mathbb{Z}$ be sets of real, complex, and integer numbers, respectively.

\section{Overview of an OTFS System \label{sec:overview}}
\begin{figure*}[tbp]
\includegraphics[clip, scale=0.45]{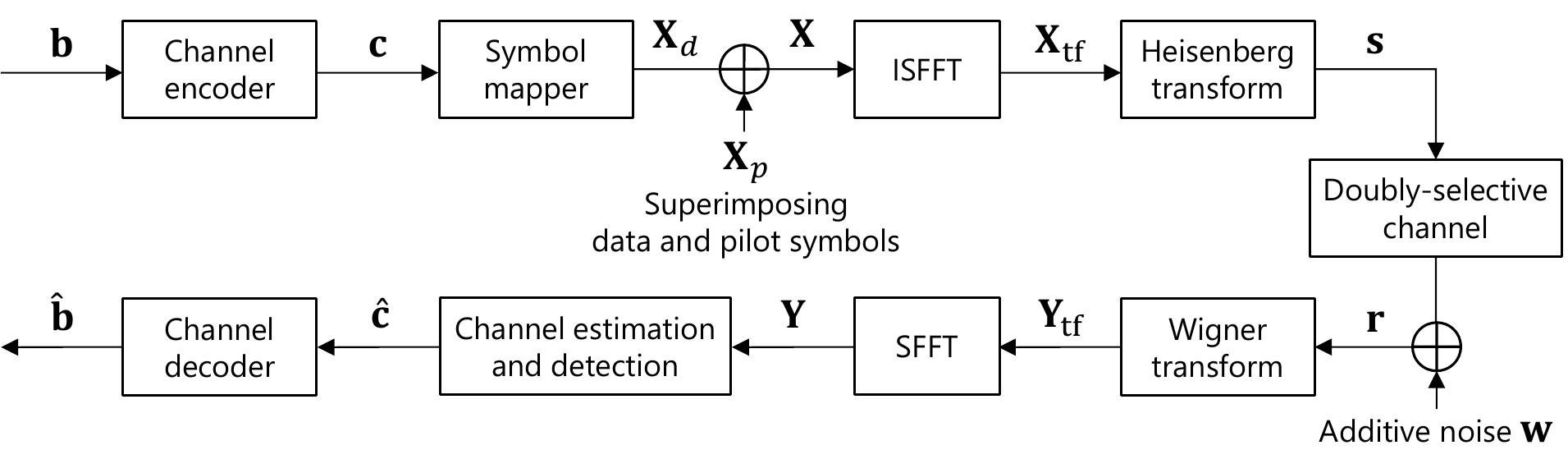}
\centering
\caption{Block diagram of the SP-based OTFS system in coded scenarios.}
\label{fig:OTFSmodel}
\end{figure*}
In this section, prior to summarize the overview of an OTFS system, we introduce some notations about the time-frequency (TF) and DD domain with reference to \cite{hong2022delaydoppler}.
Fig.~\ref{fig:OTFSmodel} illustrates the block diagram of the SP-based OTFS system in coded scenarios.
Assuming the OTFS system with $N$ timeslots and $M$ subcarriers, the discrete TF domain is defined as the $M \times N$ array of points on
$\Lambda = \{ (l \Delta f, kT), l=0,\cdots,M-1, k=0,\cdots,N-1 \}$
, where $T$ denotes the duration per one time slot and $\Delta f$ denotes the subcarrier spacing.
The discrete TF samples at points on $\Lambda$ are collected as the matrix $\Xtf[l,k] \in \mathbb{C}^{M \times N}$.

By contrast, the discrete DD domain is defined as
$\Gamma = \{ (\frac{m}{M \Delta f}, \frac{n}{NT}), m=0,\cdots,M-1, n=0,\cdots,N-1 \}$,
where $\frac{1}{M\Delta f}$ and $\frac{1}{NT}$ are the resolutions of the path delays and the Doppler shifts, respectively.
We define the discrete DD samples of the OTFS waveform at the points on $\Gamma$ as the matrix $\X[m,n] \in \mathbb{C}^{M \times N}$.

\subsection{OTFS Modulation \label{subsec:otfs_mod} \cite{hong2022delaydoppler}}
The DD domain data matrix $\X[m,n]$ is mapped to the TF domain matrix $\Xtf[l,k]$ on $\Lambda$ via inverse symplectic fast Fourier transform (ISFFT), i.e., \cite{hong2022delaydoppler}
\begin{align}
\Xtf[l, k] = \frac{1}{\sqrt{NM}}\sum_{n=0}^{N-1}\sum_{m=0}^{M-1} \X[m,n] e^{j 2\pi (\frac{nk}{N} - \frac{ml}{M})}.
\label{isfft} 
\end{align}
Next, the TF domain matrix $\Xtf[l,k]$ is converted to the continuous-time signal $s(t)$ using the transmitter pulse-shaping waveform $g_\mathrm{tx}(t)$ as \cite{hong2022delaydoppler}
\begin{align}
s(t) = \sum_{k=0}^{N-1}\sum_{l=0}^{M-1} \Xtf[l,k] g_\mathrm{tx}(t - kT) e^{j 2\pi l \Delta f (t - kT)},
\label{heisenberg} 
\end{align}
which is referred to as the Heisenberg transform.
Here, regarding the ISFFT as a combination of the $M$-point discrete Fourier transformation (DFT) of the columns and the $N$-point inverse DFT (IDFT) of the rows, the discrete time domain vector $\mathbf{s} \in \mathbb{C}^{NM \times 1}$ is obtained from \eqref{isfft} and \eqref{heisenberg} as \cite{hong2022delaydoppler}\footnote{We use the property given by $\mathrm{vec}(\mathbf{ABC}) = (\mathbf{C} \otimes \mathbf{A}) \cdot \mathrm{vec}(\mathbf{B})$ when $\mathbf{C}$ is a symmetric matrix.}
\begin{align}
\mathbf{s} = \mathrm{vec}(\Gtx \FM^{\mathrm{H}} \FM \X \FN^{\mathrm{H}}) = (\FN^{\mathrm{H}} \otimes \Gtx) \x,
\label{otfs_mod}
\end{align}
where $\FN \in \mathbb{C}^{N \times N}$ represent the $N$-point DFT matrix, the diagonal matrix $\Gtx \in \mathbb{C}^{M \times M}$ has the samples of $g_\mathrm{tx}(t)$ as its entries, and $\x \in \mathbb{C}^{NM \times 1}$ is the vectorized form of $\X$.

\subsection{OTFS Demodulation \cite{hong2022delaydoppler}}
The continuous-time received signal $r(t)$ is converted to the TF domain received matrix $\Ytf[l,k] \in \mathbb{C}^{M \times N}$ with the Wigner transform, which is given by \cite{hong2022delaydoppler}
\begin{align}
\Ytf[l,k] = \int r(t') \grx^{*}(t' - t) e^{-j 2\pi f(t' - t) }dt' |_{f=l \Delta f, t=kT},
\label{wigner}
\end{align}
where $g_\mathrm{rx}(t)$ is the receiver pulse-shaping waveform.
The DD domain received matrix $\Y[m,n] \in \mathbb{C}^{M \times N}$ is obtained from the TF domain received matrix $\Ytf[l,k]$ via SFFT as \cite{hong2022delaydoppler}
\begin{align}
\Y[m, n] = \frac{1}{\sqrt{NM}}\sum_{n=0}^{N-1}\sum_{m=0}^{M-1} \Ytf[l,k] e^{-j 2\pi (\frac{nk}{N} - \frac{ml}{M})}.
\label{sfft} 
\end{align}
As with~\eqref{otfs_mod}, the relation between the DD domain received vector $\mathbf{y} \in \mathbb{C}^{NM \times 1}$ and the discrete time domain vector $\mathbf{r} \in \mathbb{C}^{NM \times 1}$ can be expressed by \cite{hong2022delaydoppler}
\begin{align}
\mathbf{y} = \FM^\mathrm{H} \FM \Grx \mathrm{vec}^{-1}(\mathbf{r}) \FN = (\FN \otimes \Grx)\mathbf{r},
\label{otfs_demod}
\end{align}
where the diagonal matrix $\Grx \in \mathbb{C}^{M \times M}$ has the samples of $g_\mathrm{rx}(t)$ as its entries, and $\y$ is the vectorized form of $\Y$.

\subsection{Channel Model \cite{hong2022delaydoppler}}
Practically, a wireless channel with small-scale fading (also known as multipath fading) has a limited number of propagation paths with distinct delay and Doppler shift parameters, which makes the DD channel response sparsely, i.e., \cite{hong2022delaydoppler}
\begin{align}
h(\tau, \nu) = \sum_{i=1}^{P} h_i \delta(\tau - \tau_i) \delta(\nu - \nu_i),
\label{chan_response}
\end{align}
where $P$ denotes the number of propagation paths and $h_i \in \mathbb{C}, \tau_i \in \mathbb{R}_{\geq0}, \nu_i \in \mathbb{R} \ (i=1,2,\cdots,P)$ are the path gain, the delay shift, and the Doppler shift corresponding to the $i$-th path, respectively.

Assuming the actual delay shift $\tau_i$ and the Doppler shift $\nu_i$ as the integer multiples of the resolution $(\frac{1}{M \Delta f}, \frac{1}{NT})$ on the DD domain grid $\Gamma$, the normalized delay and Doppler shift $l_i \in \mathbb{Z}_{\geq0}, k_i \in \mathbb{Z}$ are given by \cite{hong2022delaydoppler}
\begin{align}
\tau_i &= \frac{l_i}{M \Delta f} \leq \tau_\mathrm{max} = \frac{\lmax}{M \Delta f} \quad \mathrm{and} \label{li} \\ 
\nu_i &= \frac{k_i}{NT} \ \mathrm{with} \ |\nu_i| \leq \nu_\mathrm{max} = \frac{\kmax}{NT},
\label{ki}
\end{align}
where $l_\mathrm{max}, k_\mathrm{max} \in \mathbb{Z}_{\geq0}$ are the normalized taps associated with the delay-Doppler spread $\tau_\mathrm{max}, \nu_\mathrm{max} \in \mathbb{R}_{\geq0}$, which correspond to the propagation distance and relative velocity of the reflectors.

\subsection{System Model \cite{raviteja2019practical}}
The time domain input-output relation between the continuous signals $s(t)$ and $r(t)$ can be expressed by \cite{raviteja2019practical}
\begin{align}
r(t) = \iint h(\tau, \nu) s(t - \tau) e^{j2\pi \nu (t - \tau)} d\tau d\nu + w(t),
\label{r_continuous}
\end{align}
where $w(t)$ denotes the additive white Gaussian noise (AWGN) term.
From \eqref{chan_response} and \eqref{r_continuous}, after the cyclic prefix (CP) removal, the discrete received signal $\mathbf{r} = \{ r(n)\}_{n=0}^{NM-1}$ is sampled at the rate $f_s = M \Delta f$ as \cite{raviteja2019practical}
\begin{align}
r(n) = \sum_{i=1}^{P} h_i e^{j2\pi \frac{k_i(n - l_i)}{NM}} s([n - l_i]_{NM}) + w(n).
\label{r_discrete}
\end{align}
We can rewritten \eqref{r_discrete} in the vectorized form, i.e., \cite{raviteja2019practical}
\begin{align}
\mathbf{r} = \mathbf{Gs} + \mathbf{w}.
\label{r_time}
\end{align}
The time domain channel matrix $\mathbf{G} \in \mathbb{C}^{NM \times NM}$ is generated from three parameters $h_i, l_i$ and $k_i$ as \cite{raviteja2019practical}
\begin{align}
\mathbf{G} = \sum_{i=1}^{P} h_i \mathbf{\Pi}^{l_i} \mathbf{\Delta}^{k_i},
\label{H_time}
\end{align}
where $\mathbf{\Pi}$ is the permutation matrix to perform a forward cyclic shift, given by \cite{raviteja2019practical}
\begin{align}
\mathbf{\Pi} = 
\begin{bmatrix}
0      & \cdots & 0      & 1      \\
1      & \ddots & 0      & 0      \\
\vdots & \ddots & \ddots & \vdots \\
0      & \cdots & 1      & 0
\end{bmatrix} \in \{0, 1\}^{NM \times NM}
\label{Pi}
\end{align}
and $\mathbf{\Delta} \in \mathbb{C}^{NM \times NM}$ is the diagonal matrix with the phase shift term $z = e^{\frac{j2\pi}{NM}}$, given by \cite{raviteja2019practical}
\begin{align}
\mathbf{\Delta} = \mathrm{diag}[z^0, z^1, \cdots, z^{NM-1}].
\label{Delta}
\end{align}
The matrices $\mathbf{\Pi}^{l_i}$ and $\mathbf{\Delta}^{k_i}$ model the delay-Doppler shifts associated with $i$-th path.

Substituting \eqref{otfs_mod} and \eqref{otfs_demod} into \eqref{r_time}, the DD domain input-output relation in vectorized form is given by \cite{raviteja2019practical}
\begin{align}
\mathbf{y} &= \underbrace{(\FN \otimes \Grx) \mathbf{G} (\FN^{\mathrm{H}} \otimes \Gtx)}_{\mathbf{H}} \x + \underbrace{(\FN \otimes \Grx)\mathbf{w}}_{\tilde{\mathbf{w}}}.
\label{y_dd}
\end{align}
where $\H$ denotes the DD domain channel matrix and $\tilde{\mathbf{w}}$ is the effective noise vector.\footnote{
Since we have the same assumption as \cite{raviteja2019practical}, the rectangular waveforms, the diagonal matrices $\Gtx$ and $\Grx$ are equivalent to an $M \times M$ identity matrix $\IM$, resulting in the same statistical properties of $\mathbf{w}$ and $\tilde{\mathbf{w}}$.
}

\section{OTFS Channel Estimation Schemes \label{sec:est}}
\begin{figure*}[tbp]
	\centering
 \subcapcentertrue
	\subfigure[EP-based scheme \cite{raviteja2019embedded}.]{
		\includegraphics[width=0.31\linewidth]{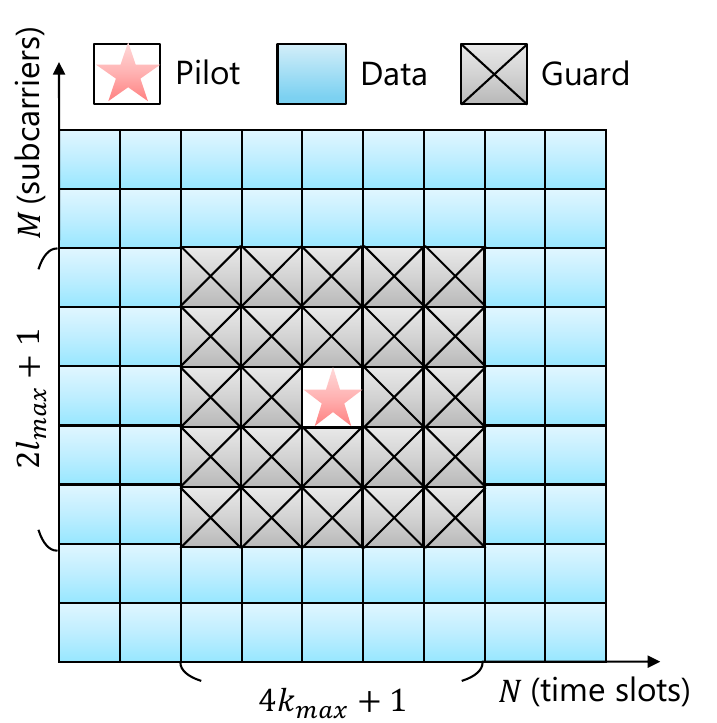}
        \label{fig:EP}
	}
	\subfigure[single SP-based scheme \cite{yuan2021dataaided}.]{
		\includegraphics[width=0.31\linewidth]{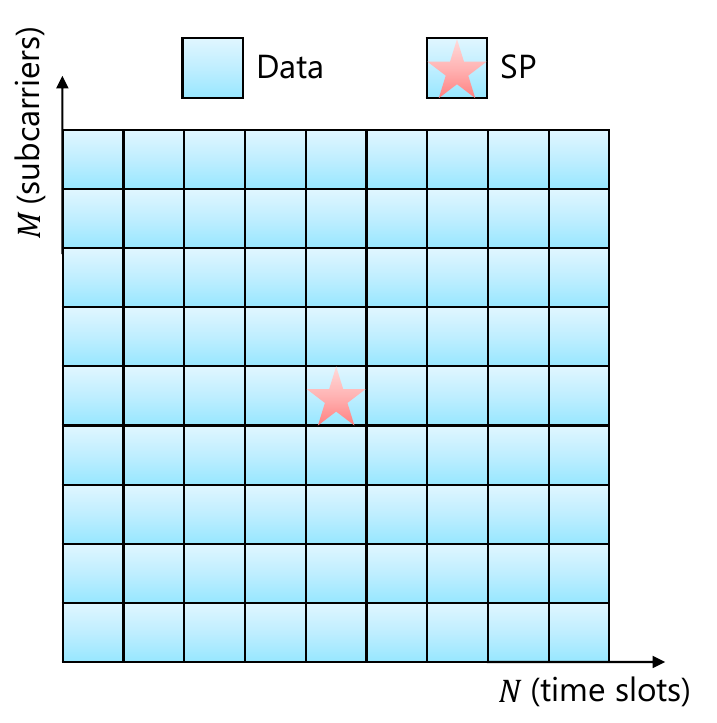}
        \label{fig:singleSP}
	}
    \subfigure[multiple SP-based scheme.]{
		\includegraphics[width=0.31\linewidth]{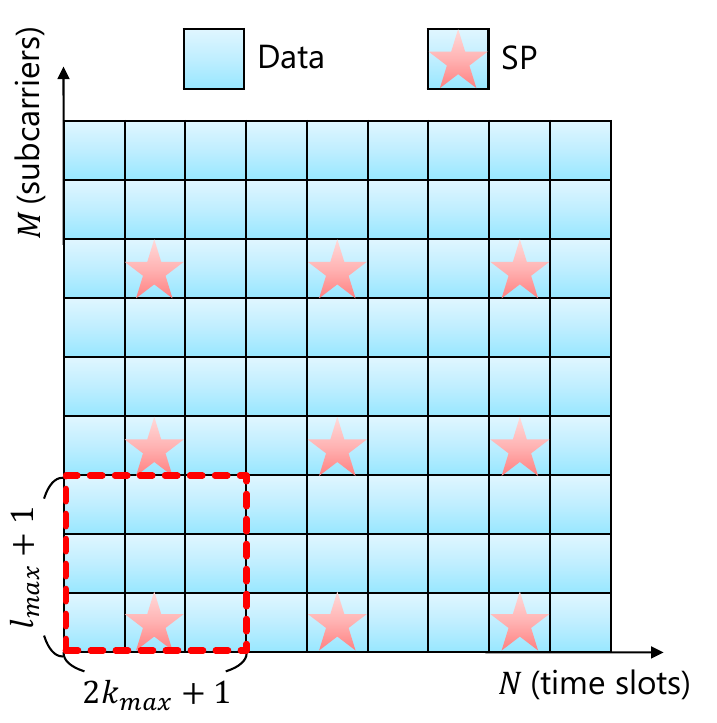}
        \label{fig:multipleSP}
	}
	\caption{Transmission frame structures for the three OTFS channel estimation schemes considered in this paper.\label{fig:frames}}
\end{figure*}

In this section, three OTFS channel estimation schemes: EP-based \cite{raviteja2019embedded}, single SP-based \cite{yuan2021dataaided}, and our proposed multiple SP-based schemes, are reviewed and compared in a fair manner.
Each frame structure is exemplified in Fig.~\ref{fig:frames}.
Note that each has a common approach using a threshold to distinguish data and pilot symbols.

There exists other SP-based approaches, such as the SP-full \cite{mishra2022otfs}, SP-sparse \cite{jesbin2023sparse}, SP-scattered \cite{liu2023low}, SP delay-wise (SP-DW) \cite{chen2023delaywise} schemes.
But, the SP-full scheme \cite{mishra2022otfs} is unable to estimate delay and Doppler shifts $(l_i, k_i)$, and has to be used together with the EP-based scheme.
The others \cite{jesbin2023sparse,liu2023low,chen2023delaywise} mainly focus on improving PAPR and rely on different channel estimators,  resulting in unfair comparisons in terms of the effective transmission power.
Thus, in order to evaluate the effects of increased number of SPs, we consider the multiple SP-based scheme of Fig.~\ref{fig:multipleSP} that can maximize the number of SPs.

\subsection{EP-based Scheme}
In the EP-based scheme, the DD domain samples matrix $\X_\mathrm{EP} \in \mathbb{C}^{M \times N}$ consists of data symbols $x_d[m, n] \in \mathbb{C}$, one pilot symbol $x_p \in \mathbb{C}$, and a guard space, which is given by
\begin{align}
\X_\mathrm{EP}[m, n] = 
    \begin{cases}
    x_p & m = m_p, \ n = n_p, \\
    0 &
    \begin{cases}
        m_p - l_\mathrm{max} \leq m \leq  m_p + l_\mathrm{max}, \\
        n_p - 2k_\mathrm{max} \leq n \leq  n_p + 2k_\mathrm{max},
    \end{cases}
    \\
    x_d[m, n] & \mathrm{otherwise}.
    \end{cases}
\label{EP_frame}
\end{align}
As given in \eqref{EP_frame} and Fig.~\ref{fig:EP}, $(2\lmax + 1)(4\kmax + 1)$ symbols are allocated for the guard space to avoid the data-pilot interference in the EP-based scheme, which realizes high channel estimation accuracy at the cost of low transmission rate.
Since the amount of the guard space depends on the delay-Doppler shift parameters $\lmax$ and $\kmax$, the transmission rate decreases correspondingly, which becomes a serious issue in high-mobility scenarios.

\subsection{SP-based Scheme}
Since the SP-based scheme superimposes data and pilot symbols in the same slot to increase the transmission rate, the DD domain samples matrix $\X_\mathrm{SP} \in \mathbb{C}^{M \times N}$ is given by
\begin{align}
\X_\mathrm{SP}[m, n] &= 
    \begin{cases}
    x_p + x_d[m, n] &
        \begin{cases}
            m = m_{p_j}, \ n = n_{p_j}, \\
            j = 1,2, \cdots, \Np,
        \end{cases} \\
    x_d[m, n] & \mathrm{otherwise},
    \end{cases}
\label{SP_frame}
\end{align}
where $m_{p_j}, n_{p_j}$ denote the indices of the $j$-th pilot symbol on the DD domain and $N_p$ is the number of pilot symbols per frame.
In \eqref{SP_frame}, the case with $N_p=1$ corresponds to the single SP-based scheme \cite{yuan2021dataaided} and the case with $N_p \neq 1$ corresponds to proposed multiple SP-based scheme.

The multiple SP-based scheme has the constraint on the pilot placement to avoid inter-pilot interference, which determines the maximum number of assignable SPs.
Specifically, multiple SPs have to be apart from each other at least $2\kmax+1$ slots on the $N$ axis and $\lmax+1$ slots on the $M$ axis.
For example, assuming $(N, M) = (9, 9)$ and $(\lmax, \kmax) = (2, 1)$, the maximum number of SPs is $9$ with a pilot arrangement given by $(l_p, k_p) = \{ (0, 1), (0, 4), (0, 7), (3, 1), (3, 4), (3, 7), (6, 1), (6, 4), (6, 7)\}$, which is illustrated in Fig.~\ref{fig:multipleSP}.

\subsection{Common Channel Estimator \label{subsec:estimator}}
Each scheme has almost the same channel estimation process despite of the different frame structures.
Ignoring the noise term, we focus on the input-output relation between the data symbols $\X$ and received symbols $\Y$ given by \cite{hong2022delaydoppler}
\begin{align}
\Y[m, n] = \sum_{i=1}^{P} h_i z^{k_i (m - l_i)} \X[[m - l_i]_M, [n - k_i]_N].
\label{I/O_DDdomain}
\end{align}
Substituting the indices $m = m_{p_j} + l_i,~n = n_{p_j} + k_i$ into \eqref{I/O_DDdomain}, the received pilot symbol associated with the $i$-th path and $j$-th pilot symbol can be rewritten as
\begin{align}
\Y[m_{p_j} + l_i, n_{p_j} + k_i] \approx h_i z^{k_i m_{p_j}} \X[m_{p_j}, n_{p_j}] \nonumber \\
=\begin{cases}
h_i z^{k_i m_{p_j}} x_p, & \textrm{(EP)} \\
h_i z^{k_i m_{p_j}} (x_p + x_d[m_{p_j}, n_{p_j}]). & \textrm{(SP)}
\end{cases}
\label{I/O_DDdomain_rewrite}
\end{align}
The received pilot symbols $\Y[m_{p_j} + l_i, n_{p_j} + k_i]$ are disturbed by the noise and data-pilot interference terms, causing channel estimation errors.
From \eqref{I/O_DDdomain_rewrite}, the estimated channel gain $\hat{h}_i \in \mathbb{C}$ can be obtained by
\begin{align}
\hat{h}_i = \frac{1}{\Np} \bigg( \sum_{j=1}^{\Np} \frac{\Y[m_{p_j} + \hat{l}_i, n_{p_j} + \hat{k}_i]}{z^{\hat{k}_i m_{p_j}}} \bigg) / x_p,
\label{hi_hat}
\end{align}
where $\hat{l}_i \in \mathbb{Z}_{\geq0}, \hat{k}_i \in \mathbb{Z}$ denote the estimated delay shift and Doppler shift, respectively.
In \eqref{hi_hat}, an average of the received SPs is calculated, which realizes the channel estimation and interference cancellation simultaneously, resulting in accurate estimation of $h_i$.

Next, we describe how to estimate the normalized delay-Doppler shifts $\hat{l}_i, \hat{k}_i$ used in \eqref{hi_hat}.
Let $b[l, k] \in \{0, 1\}$ denote whether a path with delay shift $l$ and Doppler shift $k$ exists or not, the presence of each path can be distinguished by comparing the received signal power to the specific threshold $\gamma$, i.e., 
\begin{align}
b[l, k] = 
    \begin{cases}
    1, & \sum_{j=1}^{\Np} |\Y[m_{p_j} + l, n_{p_j} + k]| \geq \gamma, \\
    0, & \mathrm{otherwise}.
    \end{cases}
\label{li_ki_hat}
\end{align}
In \eqref{li_ki_hat}, the estimated delay-Doppler shifts $\hat{l}_i$ and $\hat{k}_i$ are obtained when we find the indices $(l, k)$ which satisfies $b[l, k] = 1$, and the number of paths is $\hat{P} = \sum_{l} \sum_{k} b[l, k]$.

The threshold $\gamma$ of \eqref{li_ki_hat} depends on the specific frame structure.
When using an inappropriate threshold, it becomes difficult to accurately estimate the delay-Doppler shifts $\hat{l}_i$ and $\hat{k}_i$, leading to channel estimation errors.
The causes of such errors are classified into two cases:
\begin{enumerate}
    \item A smaller threshold makes the power of the received interference terms more likely to exceed the threshold, resulting in the detection of non-existing paths. In \eqref{hi_hat}, the channel gain $\hat{h}_i$ is estimated by dividing the received symbol $\Y[m_{p_j} + \hat{l}_i, n_{p_j} + \hat{k}_i]$ by the pilot symbol $x_p$. Since the received symbol $\Y$ composed of interference and noise has lower power than the pilot symbol $x_p$, its impact on the channel estimation accuracy is negligible.
    \item A larger threshold makes the power of the received pilot symbols more likely to fall below the threshold, resulting in the failure to detect existing paths. Assuming a channel with a uniform power profile, missing even one path causes a more serious performance degradation in terms of channel estimation accuracy compared to the first case.
\end{enumerate}
Based on these characteristics, we need to determine a proper threshold.

The EP-based scheme has a noise-dependent threshold given by \cite{raviteja2019embedded}
\begin{align}
\gamma_\mathrm{EP} &= 3 \sqrt{\sigma^2} \label{threshold_EP},
\end{align}
where $\sigma^2$ is the noise variance.
Similarly, the SP-based scheme has a threshold depending on the noise and interference terms, i.e.,
\begin{align}
\gamma_\mathrm{SP}^{(0)} &= 3 \sqrt{\Np (\sigma^2 + \sigma_d^2)}, \label{threshold_SP}
\end{align}
where $\sigma_d^2$ is the average power of data symbols.
The SP-based scheme employs the threshold $\gamma_\mathrm{SP}^{(0)}$ in the initial stage of an iterative interference cancellation algorithm, which is composed of channel estimation and symbol detection.
In the subsequent iteration stages, the threshold should be updated to a smaller value according to the remaining interference terms.
But, deriving the optimal updated threshold is a challenging task, as the progress of the interference cancellation depends on the accuracy of the estimated data symbols and channel matrices for each trial.
Thus, after the initial stage of the algorithm, the threshold $\gamma_\mathrm{SP}^{(0)}$ is given by
\begin{align}
\gamma_\mathrm{SP}^{(r)} &= 3 \sqrt{\Np \cdot \sigma^2}
\label{threshold_iter}
\end{align}
for $r \geq 1$, where $r$ is an iteration index.

The estimated channel matrix $\Hhat \in \mathbb{C}^{NM \times NM}$ is finally generated from these channel parameters $\hat{h}_i, \hat{l}_i, \hat{k}_i$ and $\hat{P}$.

\section{Proposed Algorithm for Coded Scenarios\label{sec:algo}}
In this section, we propose the iterative algorithm which eliminates data-pilot interference precisely, utilizing the error correction capability of the channel coding.
We also describe key techniques used in the proposed algorithm: symbol detection, channel coding, and interference cancellation process.
In addition, the computational complexity analysis is performed for the EP-based and SP-based schemes.

\subsection{Symbol Detection}
We apply the MP detection algorithm \cite{raviteja2018interference}, which offers better performance and lower complexity than the linear minimum mean square error detection due to the sparsity of channel matrix.
Although outputs of the MP detection $\hat{\x} \in \mathbb{C}^{NM \times 1}$ are typically given by hard decisions on the received symbols, the channel decoder requires the LLRs for every bit.
Therefore, we focus on the probability mass function (PMF) $\mathbf{P} \in \mathbb{R}^{NM \times L}$, corresponding to the probability of the information symbols on the constellation points of size $L$, which is updated in the process of the MP detection.
In order to obtain the LLR information, the matrix $\mathbf{Z} \in \mathbb{R}^{NM \times L}$, which denotes the logits for every symbol, is calculated by
\begin{align}
\mathbf{Z}[i, j] = \mathrm{logit}(\mathbf{P}[i, j]) = \log \bigg( \frac{\mathbf{P}[i,j]}{1 - \mathbf{P}[i, j]} \bigg)
\label{logits}
\end{align}
for $i = 0,1,\cdots,NM-1$ and $j = 0,1,\cdots,L-1$.
The LLR vector $\mathbf{L} \in \mathbb{R}^{R_c}$ is obtained from the matrix $\mathbf{Z}$ as
\begin{align}
\mathbf{L}[k] = \log \bigg( \frac{\mathrm{Pr} (\mathbf{c}[k]=0 | \mathbf{Z})}{\mathrm{Pr} (\mathbf{c}[k]=1 | \mathbf{Z})} \bigg)
\label{LLR}
\end{align}
for $k = 0,1,\cdots,R_c-1$, where $R_c$ denotes the number of coded bits and $\mathbf{c} \in \{0, 1\}^{R_c}$ is the coded bit sequence.
From \eqref{logits} and \eqref{LLR}, the LLR vector $\mathbf{L}$ for every bit is passed to the decoder as input information.

\subsection{Channel Coding}
The channel encoder has the number of the coded bits $R_c$ and information bits $R_b$ as input parameters, and the coding rate $r_c$ is defined as $r_c = R_b/R_c$.
Then, the information bit sequence $\mathbf{b} \in \{ 0, 1 \}^{R_b}$ is encoded to the coded bit sequence $\mathbf{c}$ via the encoder.

By contrast, the channel decoder is designed with the iterative belief propagation (BP) algorithm, which requires the number of iterations $I_\mathrm{LDPC}$ as an input parameter.
The BP decoder performs error correction using the input LLR vector $\mathbf{L}$, and provides the estimated bit sequence $\hat{\mathbf{b}} \in {\{ 0,1 \}^{R_b}}$ and the updated LLR vector $\hat{\mathbf{L}} \in \mathbb{R}^{R_c}$ as outputs.
At the end of the iterative algorithm, the hard-decided bit sequence $\hat{\mathbf{b}}$ is provided.
Considering the interference cancellation process, the decoder provides the soft-decision output, i.e., the updated LLR vector $\hat{\mathbf{L}}$.

Note that we implement 5G compliant LDPC codes based on Sionna\footnote{\url{https://nvlabs.github.io/sionna/}}, which is an open source library for simulating the physical layer of wireless and optical communication systems.

\subsection{Proposed Iterative Algorithm}
\begin{figure*}[tbp]
    \centering
    \includegraphics[clip, scale=0.45]{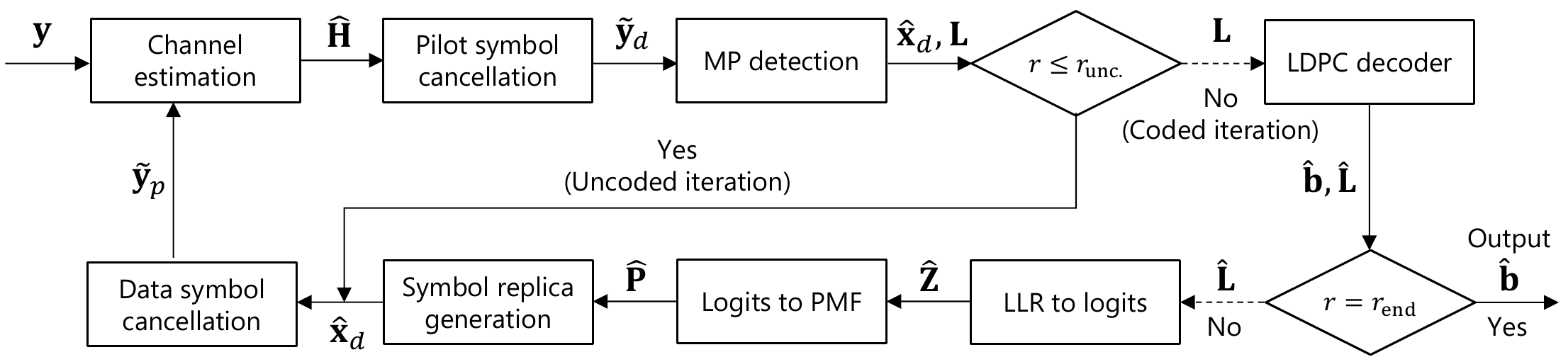}
    \caption{Proposed iterative algorithm with channel coding.}
    \label{fig:iterative_model}
\end{figure*}
\begin{algorithm}[tb]
    \caption{Proposed iterative algorithm.}
    \begin{algorithmic}[1]
        \renewcommand{\algorithmicrequire}{\textbf{Input:}}
        \renewcommand{\algorithmicensure}{\textbf{Output:}}
        \REQUIRE Received symbols $\mathbf{y}$ and known pilot symbols $\x_p$ % Input
        \STATE \textbf{Initialization:} Iteration index is set to $r=1$.
        \REPEAT
            \STATE \textbf{Channel estimation:} Channel parameters $\hat{P}, \hat{h}_i, \hat{l}_i,$ and $\hat{k}_i$ are estimated by \eqref{hi_hat} and \eqref{li_ki_hat}. The estimated channel matrix $\hat{\H}$ is obtained by \eqref{H_time}--\eqref{y_dd}.
            \STATE \textbf{Pilot symbol cancellation:} The received data symbols $\tilde{\mathbf{y}}_d$ are extracted by \eqref{yd_tilde}.
            \STATE \textbf{MP detection:} The MP detection is performed to obtain the estimated data symbols $\hat{\x}_d$ and LLR vector $\mathbf{L}$ by \eqref{logits} and \eqref{LLR}.
            \IF{$r \leq r_\mathrm{unc.}$}
                \STATE The uncoded iteration is applied and the estimated data symbols $\hat{\x}_d$ are obtained as outputs of the MP detector.
            \ELSE
                \STATE The coded iteration is applied and the LLR vector $\mathbf{L}$ is obtained as an output of the MP detector.
                \STATE \textbf{LDPC decoder:} The LDPC decoder performs error correction to obtain the estimated bit sequence $\hat{\mathbf{b}}$ and updated LLR vector $\hat{\mathbf{L}}$.
                \STATE \textbf{LLR to logits:} The LLR vector $\hat{\mathbf{L}}$ is converted to the updated logits matrix $\hat{\mathbf{Z}}$ by \eqref{LLR_to_logits}.
                \STATE \textbf{Logits to PMF:} The logits matrix $\hat{\mathbf{Z}}$ is converted to the updated PMF matrix $\hat{\mathbf{P}}$ by \eqref{Logits_to_PMF}.
                \STATE \textbf{Symbol replica generation:} As described in \eqref{replica_generation}, the soft-decision symbol replicas $\hat{\mathbf{x}}_d$ are generated from the PMF matrix $\hat{\mathbf{P}}$.
            \ENDIF
            \STATE \textbf{Data symbol cancellation:} The received pilot symbols $\tilde{\mathbf{y}}_p$ are extracted by \eqref{yp_tilde}, which is utilized for the next channel estimation instead of $\y$.
            \STATE $r \gets r + 1$.
        \UNTIL{A stopping criteria is satisfied.}
        \ENSURE Estimated bit sequence $\hat{\mathbf{b}}$ obtained in Line~10  % Output
    \end{algorithmic}
    \label{algorithm}
\end{algorithm}

In the SP-based scheme, the received symbols contain inevitable data-pilot interference, resulting in poor channel estimation accuracy.
Thus, the most of SP-based schemes perform the interference cancellation for the iterative channel estimation, using the estimated channel $\Hhat$ and data symbols $\hat{\mathbf{x}}_d \in \mathbb{C}^{NM \times 1}$, i.e.,
\begin{align}
\tilde{\mathbf{y}}_p = \mathbf{y} - \Hhat \hat{\mathbf{x}}_d,
\label{yp_tilde}
\end{align}
which extracts the received pilot symbols $\tilde{\mathbf{y}}_p \in \mathbb{C}^{NM \times 1}$ from the received symbols $\y$.
By contrary to \eqref{yp_tilde}, to perform the iterative symbol detection, the received data symbols $\tilde{\mathbf{y}}_d \in \mathbb{C}^{NM \times 1}$ are extracted from the received symbols $\mathbf{y}$ as
\begin{align}
\tilde{\mathbf{y}}_d = \mathbf{y} - \Hhat \mathbf{x}_p,
\label{yd_tilde}
\end{align}
where $\mathbf{x}_p \in \mathbb{C}^{NM \times 1}$ denotes the pilot symbols known a priori.
Although the refined symbols $\tilde{\y}_p$ and $\tilde{\y}_d$ are used for the iterative channel estimation and symbol detection, remained interference induces the error floor in the high SNR region, which is an open issue in the SP-based scheme.

Against this challenge, to further improve channel estimation accuracy, we propose the iterative algorithm illustrated in Fig.~\ref{fig:iterative_model}.
It is also summarized in Algorithm~$\ref{algorithm}$.
The proposed algorithm utilizes the LLR information to generate replicas of the data symbols, which is motivated by a prior study in a physical layer security \cite{hama2023serially}.

As opposed to the conventional algorithm in \cite{yuan2021dataaided}, the proposed one receives the LLR vector $\mathbf{L}$ from the MP detector through \eqref{logits} and \eqref{LLR}, instead of the estimated data symbols $\hat{\mathbf{x}}_d$.
The updated LLR vector $\hat{\mathbf{L}}$ is obtained through the LDPC decoder, and it is converted to the updated logits matrix $\hat{\mathbf{Z}} \in \mathbb{R}^{NM \times L}$.
Here, the logit $\hat{\mathbf{Z}}[i, j]$ is given by a log probability for $j$-th symbol point $x_j \in \mathbb{C}$ on a constellation set $\mathcal{X}$ of size $L$, i.e.,
\begin{align}
\hat{\mathbf{Z}}[i, j] &= \log (\mathrm{Pr}(x_j \in \mathcal{X}| \hat{\mathbf{L}}_i)) \nonumber \\
&= \sum_{k=0}^{K-1} \log \bigg(\frac{1}{1 + \exp{(-\hat{\mathbf{L}}_i[k] b(x_j)_k)}}\bigg),
\label{LLR_to_logits}
\end{align}
for $k = 0,1,\cdots,\log_2{L}-1$, where $\hat{\mathbf{L}}_i \in \mathbb{R}^{\log_2{L}}$ is the LLR information vector corresponding to each symbol, and $b(x_j)_k$ represents the $k$-th bit label of $x_j$, with $0$ replaced by $-1$.
The logits matrix $\hat{\mathbf{Z}}$ is converted to the updated PMF matrix $\hat{\mathbf{P}} \in \mathbb{R}^{NM \times L}$, which can be calculated by
\begin{align}
\hat{\mathbf{P}}[i, j] = \frac{1}{ 1 + \exp{ (- \hat{\mathbf{Z}}[i, j]) } }.
\label{Logits_to_PMF}
\end{align}
The soft-decision symbol replicas $\hat{\mathbf{x}}_d$ are generated by computing a sum of the weighted constellation points $x_j$ according to its probability distribution, i.e.,
\begin{align}
\hat{\mathbf{x}}_d[i] = \sum_{j=0}^{L-1} \hat{\mathbf{P}}[i, j] \cdot x_j,
\label{replica_generation}
\end{align}
which are finally utilized to the data symbol cancellation given by \eqref{yp_tilde}.

As described in Algorithm~\ref{algorithm}, there exist two iterative processes, uncoded and coded iterations.
On one hand, the uncoded iteration process, which is identical to the conventinal algorithm in \cite{yuan2021dataaided}, has low computational complexity but it suffers remained interference in the high SNR region. 
On the other hand, the coded iteration process eliminates data-pilot interference precisely, but it has increased computational complexity because of additional operations such as the decoding and symbol replica generation.
Here, let $\runc$ and $\rcod$ be the number of uncoded and coded iterations, respectively, which are thresholds for switching two processes.
The total number of iterations is given by $\rend = \runc + \rcod$ and is a stopping criteria for the iterative algorithm.
Considering a trade-off between the channel estimation accuracy and computational complexity, we can adjust these parameters $\runc$, $\rcod$, and $\rend$ so as to implement the proposed algorithm in a flexible manner.
Details are discussed in Section~\ref{sec:simu}.

\subsection{Computational Complexity}
The computational complexities of the EP-based \cite{raviteja2019embedded}, single SP-based \cite{yuan2021dataaided}, and multiple SP-based schemes are summarized in Table~\ref{tab:complexity}.
Here, since wall-clock runtime is highly dependent on hardware and software implementations, we evaluated the overall computational complexity by counting the number of real-valued multiplications.
Here, $I_\mathrm{MP}$ denotes the number of iteration for the MP detection.
From \eqref{hi_hat} and \eqref{li_ki_hat}, channel estimation processes have the complexity depending on the channel parameters $P, \lmax, \kmax$, and the number of pilots $N_p$.
From \eqref{yp_tilde} and \eqref{yd_tilde}, the interference cancellation processes are expected to have quadratic complexities due to matrix multiplications, but actual complexities are alleviated because the channel matrix $\H$ and pilot symbols $\mathbf{x}_p$ have only $P$ and $N_p$ nonzero elements in each column, respectively.
According to \cite{raviteja2018interference}, the MP detector has a linear complexity in $N$ and $M$.
The BP-based LDPC decoder has a complexity of $\O(J R_c I_\mathrm{LDPC}) \approx \O(R_c I_\mathrm{LDPC})$, assuming a sparse parity-check matrix which contains $J$ ones in each column, i.e., $R_c \gg J$ \cite{fossorier1999reduced}.
From \eqref{LLR_to_logits}, the LLR information for $\log_2{L}$ bits is converted to the logits information for $L$ constellation points, resulting in a complexity of $O(N M L \log_2{L})$.
From \eqref{Logits_to_PMF} and \eqref{replica_generation}, the conversion of the logits to PMF and the symbol replica generation have complexities of $\O(NML)$.

\begin{table*}[tb]
    \caption{Computational complexities associated with the EP-based \cite{raviteja2019embedded}, single SP-based \cite{yuan2021dataaided}, and multiple SP-based schemes.}
    \centering
    \begin{tabular}{|c|c|c|c|}
        \hline
         & EP-based scheme & Single SP-based scheme & Multiple SP-based scheme \\
        \hline
        Channel estimation & $6P + (\lmax+1)(2\kmax+1)=\O(P + \lmax\kmax)$ & $\O(\rend(P + \lmax\kmax))$ & $\O(\rend N_p(P + \lmax\kmax))$ \\
        \hline
        Pilot symbol cancellation & $\O(P + NM)$ & $\O(\rend(P + NM))$ & $\O(\rend(PN_p + NM))$ \\
        \hline
        MP detection & $\O(I_\mathrm{MP}NMPL)$ \cite{raviteja2018interference} & \multicolumn{2}{c|}{$\O(\rend I_\mathrm{MP}NMPL)$}\\
        \hline
        LDPC decoder & $\O(R_c I_\mathrm{LDPC})$ \cite{fossorier1999reduced} & \multicolumn{2}{c|}{$\O(\rcod R_c I_\mathrm{LDPC})$} \\
        \hline
        LLR to logits & -- & \multicolumn{2}{c|}{$(\rcod - 1)\O(NML\log_2 L) = \O(\rcod NML\log_2 L)$} \\
        \hline
        Logits to PMF & -- & \multicolumn{2}{c|}{$(\rcod - 1)\O(NML) = \O(\rcod NML)$}\\
        \hline
        Symbol replica generation & -- & \multicolumn{2}{c|}{$(\rcod - 1)\O(NML) = \O(\rcod NML)$} \\
        \hline
        Data symbol cancellation & -- & \multicolumn{2}{c|}{$(\rend - 1)NM(P+1) = \O(\rend NMP)$} \\
        \hline
    \end{tabular}
    \label{tab:complexity}
\end{table*}
As shown in Table~\ref{tab:complexity}, the SP-based schemes have high complexities compared to the EP-based scheme, because of the iterative processes such as the interference cancellation and symbol replica generation.
By contrary, the computational complexities of the single SP-based and multiple SP-based schemes are almost identical because the number of pilots $N_p$ only affects the channel estimation and pilot symbol cancellation processes, which are not dominant factors in the practical assumption, i.e., $N, M, R_c \gg P, \lmax,$ and $\kmax$.

\section{Simulation Results\label{sec:simu}}
In this section, we compare the EP-based \cite{raviteja2019embedded}, single SP-based \cite{yuan2021dataaided}, and proposed multiple SP-based schemes.
As a reference, the perfect CSI (PCSI) case is considered for the SP-based arrangement in order to investigate the optimality of the proposed algorithm.
In addition, we demonstrate the advantage of the proposed algorithm through numerical simulations.
Unless otherwise specified, we consider an LDPC-coded OTFS system with QPSK signaling ($L=4$) and the coding rate of $r_c=0.75$ where the number of timeslots and subcarriers are $N = M = 15$, the carrier frequency is $f_c=4~\mathrm{GHz}$, and the subcarrier spacing is $\Delta f=15~\mathrm{KHz}$.
We apply the doubly-selective fading channel with the number of paths $P=4$, where each channel gain follows i.i.d complex Gaussian distribution with uniform power profile $h_i \in \mathcal{CN}(0, 1/P)$, which follows \cite{raviteja2019otfs}.
The normalized integer delay shift $l_i$ and integer Doppler shift $k_i$ are randomly generated in the intervals $[0, \lmax]$ and $[-\kmax, \kmax]$, similar to \cite{yuan2021dataaided}, which are idealized assumptions.
Note that we set fixed maximum delay shift $\lmax=4$ and Doppler shift $\kmax=2$, which correspond to delay spread $\tau_\mathrm{max} = \lmax / (M \Delta f) = 17.8~\mathrm{\mu s}$ and Doppler spread $\nu_\mathrm{max} = \kmax / (NT) = 2.0~\mathrm{kHz}$, respectively.
Given these parameters, the maximum number of pilots is $N_p=9$ for the multiple-SP scheme, in accordance with the constraint of the pilot assignment scheme described in Section \ref{sec:est}.

\begin{table}[tb]
  \centering
  \caption{Code lengths used for the EP and SP-based schemes.}
  \label{tab:code_length}
  \begin{tabular}{|l|l|}
    \hline
    Pilot assignment scheme & Code length [bit] \\
    \hline
    EP with QPSK     & 8064 \\
    EP with 8-PSK    & 8208 \\
    EP with 16-QAM   & 8064 \\
    SP with QPSK   & 8100 \\
    SP with 16-QAM   & 8100 \\
    \hline
  \end{tabular}
\end{table}
The code lengths used for EP and SP-based schemes are listed in Table~\ref{tab:code_length}. As shown in the table, the code lengths differ slightly due to variations in the number of data symbols and information bits per frame, but this difference is negligible and does not affect the relative performance among the pilot assignment schemes.

\subsection{SNR Definition}
In the prior studies \cite{raviteja2019embedded, yuan2021dataaided}, symbol-by-symbol data and pilot SNRs are defined individually.
Since the EP and SP-based schemes have different number of data symbols per frame, the conventional definition causes unfair comparisons in terms of the effective transmission power.
Thus, we apply a proper SNR definition to perform fair comparisons between each scheme.

Firstly, an effective SNR, which considers the power of data and pilot signals together, is defined by $\mathrm{SNR} = E_s / N_0$, where $E_s$ denotes the total transmission power per frame and $N_0 = \sigma^2 \cdot NM$ is the noise variance associated with $NM$ symbols.
The total power $E_s$ is distributed to data and pilot power denoted as $E_d$ and $E_p$, which satisfies $E_s = E_p + E_d$.
Using a parameter $\alpha \in [0, 1]$, the superimposed pilot power ratio is given by $E_p / E_d = \alpha / (1 - \alpha)$.
Next, the average power of data and pilot symbols are defined by $\sigma_d^2 = E_d / N_d$ and $\sigma_p^2 = E_p / N_p$, where $N_d$ and $N_p$ denote the number of data and pilot symbols per frame.
That is, the total pilot energy $E_p$ is kept constant for all the considered schemes with different numbers of superimposed pilots $N_p$.
Finally, assuming the normalized noise variance $\sigma^2 = 1$, data and pilot symbols are generated as $|x_d[m, n]|^2 = \sigma_d^2$ and $|x_p|^2 = \sigma_p^2$.

Setting the pilot power ratio $\alpha$ is a nontrivial task.
A low $\alpha$ induces poor channel estimation accuracy due to data-pilot interference, whereas a high $\alpha$ worsens reliability because of insufficient data power. The optimal $\alpha$ depends on multiple factors, such as SNR, modulation order, and the specific pilot assignment scheme, making it challenging to derive a closed-form solution. Therefore, in this paper, $\alpha$ is explored through preliminary simulations to minimize the uncoded BER performance. The parameter search is conducted separately for the EP and SP-based schemes, as the optimal $\alpha$ differs between the two. The search range is set from $\alpha = 0.1$ to $\alpha = 0.9$, with a step size of 0.1. By adaptively determining $\alpha$ for each SNR, the data and pilot power can be allocated in such a way that each scheme achieves near-optimal throughput performance.
These simulations only need to be computed once offline, and both the time and space complexities are negligible.

\subsection{Performance Metrics}
To characterize the performance of each scheme in a comprehensive manner, we apply three metrics; reliability, channel estimation accuracy, and spectral efficiency.
Following prior studies, reliability and channel estimation accuracy are evaluated by BER and NMSE, respectively.
The NMSE of the channel matrices is defined as
\begin{align}
\mathrm{NMSE} = \frac{\mathrm{E}[\| \hat{\H} - \H \|_\mathrm{F}^2]}{\mathrm{E}[\| \H \|_\mathrm{F}^2]}.
\label{NMSE}    
\end{align}

As a performance metric for spectral efficiency, we focus on the effective throughput which depends on the block error rate (BLER), defined as \cite{shen2003effective} 
\begin{align}
\bar{T}_u \triangleq \frac{(1 - \mathrm{BLER}) N_d r_c \log_2{L}}{NM}.
\label{throughput}
\end{align}
This metric is suitable to evaluate system performance, taking into account multiple factors such as the modulation order, the coding rate, and the pilot overhead.

\subsection{Effects of the Switching Parameters $(\runc, \rcod, \rend)$}
To confirm the advantage of the proposed algorithm, we investigated the relation between the channel estimation accuracy and iteration parameters such as $\runc, \rcod$, and $\rend$.
In this section, we consider the multiple SP-based scheme with the maximum number of SPs, i.e., $N_p=9$.

\begin{figure*}[tbp]
    \centering
    \subcapcentertrue
	\subfigure[$\mathrm{SNR}=10~\mathrm{dB}$.]{
	\includegraphics[width=0.315\linewidth]{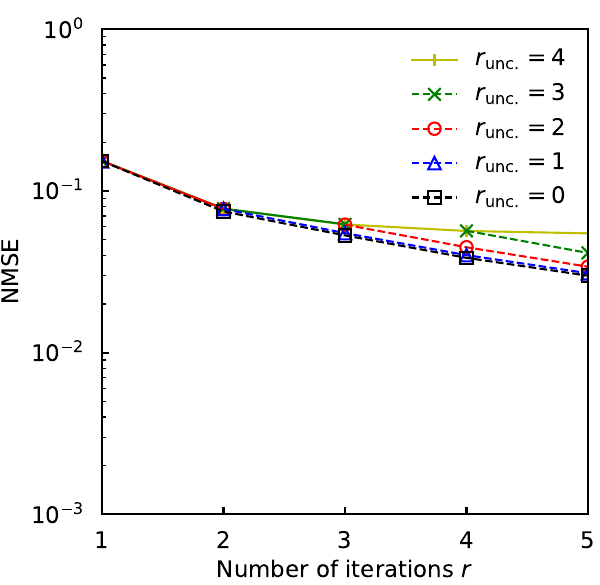}
        \label{fig:NMSE_rend_10-}
	}
        \subfigure[$\mathrm{SNR}=12.5~\mathrm{dB}$.]{
	\includegraphics[width=0.315\linewidth]{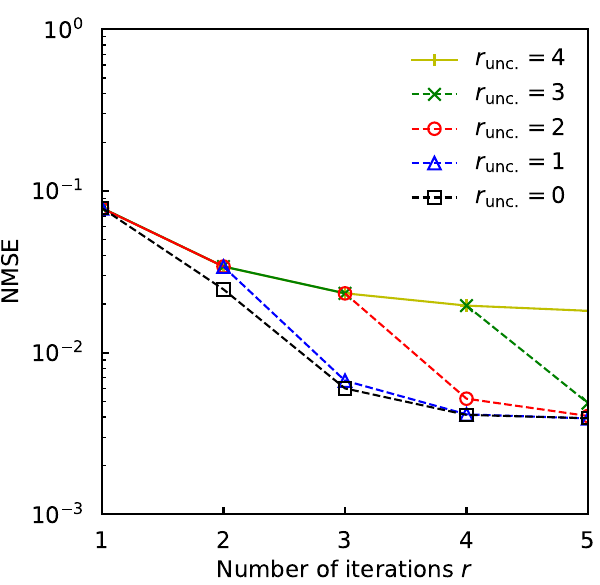}
        \label{fig:NMSE_rend_12.5}
	}
	\subfigure[$\mathrm{SNR}=15~\mathrm{dB}$.]{
	\includegraphics[width=0.315\linewidth]{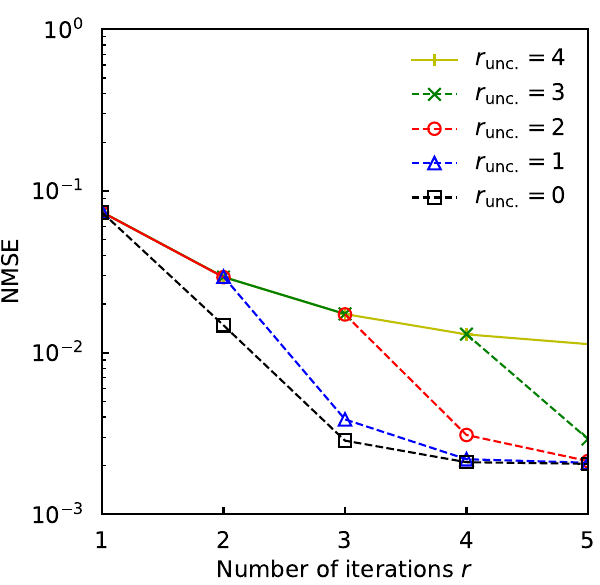}
        \label{fig:NMSE_rend_15}
	}
	\caption{NMSE comparison upon increasing the number of iterations $r$, where the stopping criteria was $\rend=5$.}
        \label{fig:NMSE_rend}
\end{figure*}
Fig.~\ref{fig:NMSE_rend} shows NMSE comparisons upon increasing the number of iteration with a fixed stopping criteria $\rend=5$ and SNR values $10$, $12.5$, and $15~\mathrm{dB}$.
Since the parameter $\runc$ denotes the number of the uncoded iterations, the case of $\runc=0$ corresponds to the proposed algorithm composed only of the coded iterations.
Contrary, in the case of $\runc=4$, it is composed of the uncoded iterations, except for the last stage of the algorithm.
As can be seen in Fig.~\ref{fig:NMSE_rend}, a smaller value of $\runc$, i.e., a larger number of coded iterations, resulted in better channel estimation accuracy.
This characteristic was significant at the high SNR.
It was also found that the proposed algorithm improved channel estimation accuracy by increasing the number of iterations $r$, but it tended to converge after a few trials.
Based on these results, a stopping criteria is set to $\rend=4$ in following simulations.

\begin{figure}[tbp]
	\centering
        \includegraphics[clip, scale=0.64]{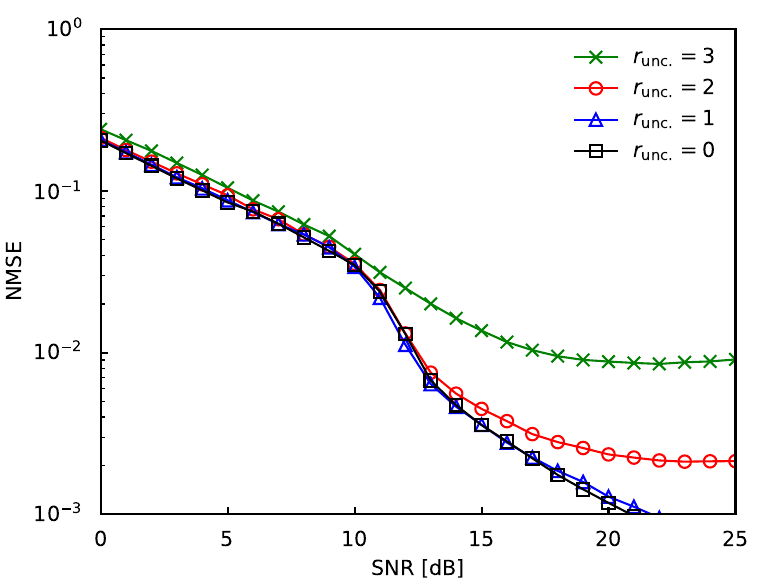}
	\caption{NMSE comparison for the multiple SP-based scheme with the number of pilots $N_p=9$, where we set iteration parameters $0 \leq \runc \leq 3$ and a fixed stopping criteria $\rend = 4$.}
    \label{fig:NMSE_runc}
\end{figure}
Fig.~\ref{fig:NMSE_runc} shows an NMSE comparison for the number of uncoded iteration $0 \leq \runc \leq 3$.
In Fig.~\ref{fig:NMSE_runc}, the superimposed pilot power ratio is fixed at $\alpha=0.5$.\footnote{
Since the channel estimation accuracy depends on the pilot power, employing an adaptive $\alpha$ leads to fluctuations in NMSE performance as SNR increases.
In order to investigate the effects of the parameters $(\runc, \rcod, \rend)$, a fixed $\alpha$ is appropriate for NMSE comparisons.
}
Note that in the case of $\runc=3$, the LLR information is not used for symbol replica generation, which is equivalent to the conventional algorithm \cite{yuan2021dataaided} in coded scenarios.
It can be observed from Fig.~\ref{fig:NMSE_runc} that the coded iteration process improved the channel estimation accuracy and its advantage was clear in the high SNR region.
For $\runc=0,1,$ and $2$, the NMSE characteristics changed depending on a specific SNR, because the accuracy of the symbol replicas depends on the error correction capability of the decoder.
These results indicate that $\runc=0$ is the best parameter setting in terms of the channel estimation accuracy.

\begin{figure}[tbp]
    \centering
    \includegraphics[clip, scale=0.64]{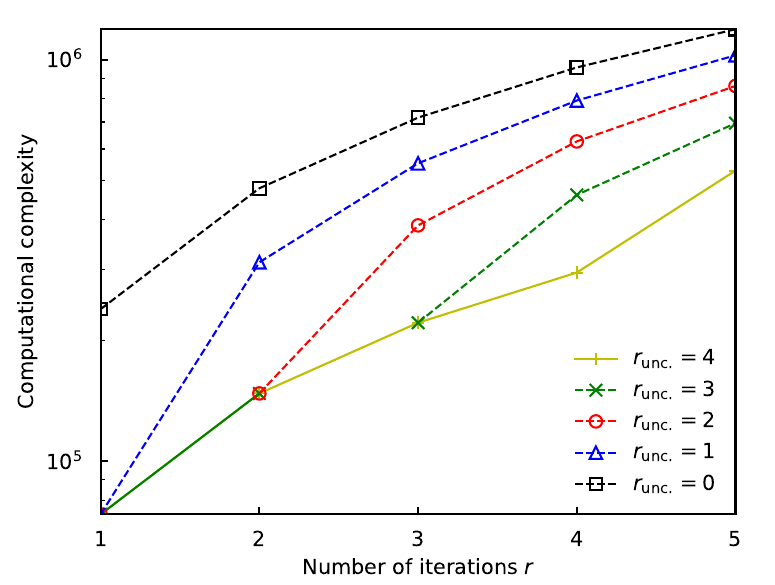}
    \caption{Computational complexity comparison upon increasing the number of uncoded iteration $\runc$, based on the complexity orders summarized in Table~\ref{tab:complexity}.}
    \label{fig:comp}
\end{figure}
In addition, Fig.~\ref{fig:comp} shows computational complexity comparison as a function of the number of iterations $r$, which is calculated from the complexity orders summarized in Table~\ref{tab:complexity}.
As expected, computational complexity increased depending on the number of iterations.
The coded iteration requires additional processes such as decoding and symbol replica generation, which result in increased complexity.

From Figs.~\ref{fig:NMSE_rend}--\ref{fig:comp}, we observed the trade-off between the channel estimation accuracy and computational complexity.
Since the iterative parameters are mutable under the condition $\runc + \rcod = \rend$, the proposed algorithm can be adjusted to achieve the desirable performance while minimizing computational complexity.

\subsection{Performance Comparison}
Through performance comparisons among the EP-based, single SP-based, and multiple SP-based schemes, we confirmed the advantage of our proposed approaches.
Before a detailed discussion, we describe the parameter settings for following simulations.
The multiple SP-based scheme allocates $N_p = 3, 6$, and $9$ pilots per frame.
For the single and multiple SP-based schemes, we employed the iterative parameters $\runc=3$ and $\rend=4$, corresponding to the conventional algorithm \cite{yuan2021dataaided}.
In addition, we employed $\runc=0$ and $\rend=4$ for the SP-based scheme with $N_p=9$, which is labeled as \textit{Iter. alg.} in the following figures.
As clarified in the previous section, these parameters realize the best interference cancellation performance with a high computational complexity trade-off.

\begin{figure}[tbp]
	\centering
        \includegraphics[clip, scale=0.64]{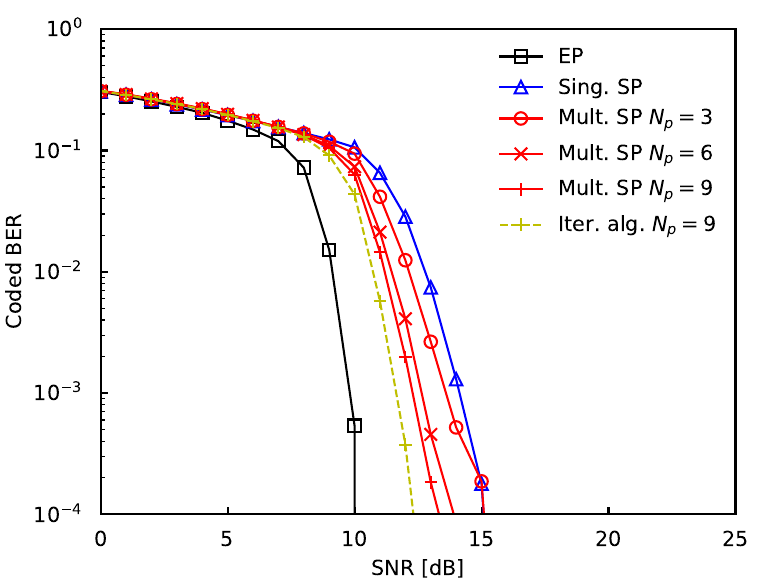}
	\caption{Coded BER comparison for the EP-based \cite{raviteja2019embedded}, single SP-based \cite{yuan2021dataaided}, and multiple SP-based schemes with the number of pilots $N_p=3,6$, and $9$.}
    \label{fig:BER}
\end{figure}
First, Fig.~\ref{fig:BER} shows a coded BER comparison.
It can be seen from Fig.~\ref{fig:BER} that the EP-based scheme achieved the best BER performance at the cost of guard space, resulting in a lower transmission rate.
Comparing the SP-based schemes with the conventional algorithm, the BER performance was improved by increasing the number of pilots $N_p$.
In addition, the BER performance was further enhanced by applying the proposed algorithm for the multiple SP-based scheme with $N_p=9$.
These results demonstrated the advantage of the proposed pilot assignment scheme and iterative algorithm.

\begin{figure}[tbp]
	\centering
        \includegraphics[clip, scale=0.64]{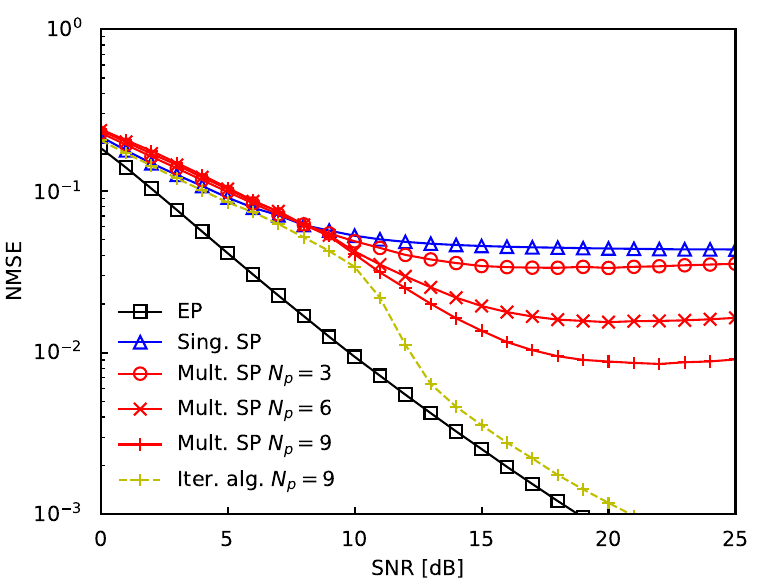}
	\caption{NMSE comparison for the EP-based \cite{raviteja2019embedded}, single SP-based \cite{yuan2021dataaided}, and multiple SP-based schemes, where the parameters were the same as those used in Fig.~\ref{fig:BER}, except for the superimposed pilot power ratio $\alpha$.}
    \label{fig:NMSE}
\end{figure}
Next, Fig.~\ref{fig:NMSE} shows an NMSE comparison for each scheme.
As in Fig.~\ref{fig:NMSE_runc}, the superimposed pilot power ratio is fixed at $\alpha = 0.5$.
As with the case of the BER comparison in Fig.~\ref{fig:BER}, it can be seen from Fig.~\ref{fig:NMSE} that the EP-based scheme achieved the best NMSE performance and the multiple SP-based scheme outperformed the single SP-based scheme.
Due to the refined symbol replicas based on the LLR information, the proposed algorithm achieved comparable NMSE performance to that of the EP-based scheme in the intermediate-to-high SNR region.

\begin{figure*}[tbp]
	\centering
 \subcapcentertrue
	\subfigure[Throughput comparison with the equal coding rate $r_c = 0.75$.]{
	\includegraphics[width=0.48\linewidth]{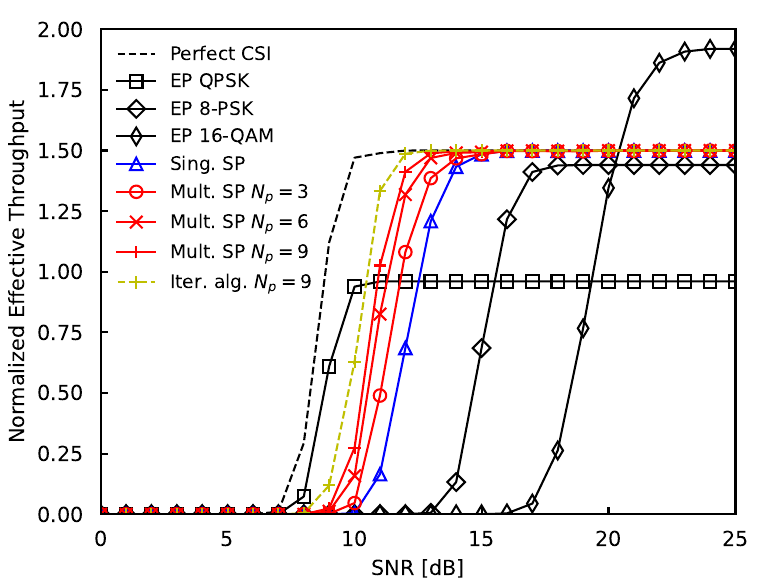}
        \label{fig:T_coderate}
	}
	\subfigure[Throughput comparison with the equal transmission rate $1.5~\mathrm{bit/sym}$.]{
	\includegraphics[width=0.48\linewidth]{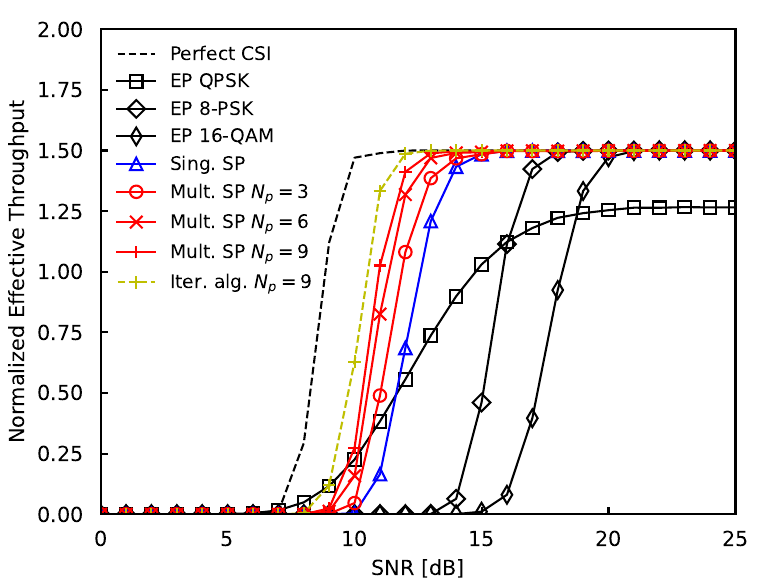}
        \label{fig:T_transmissionrate}
	}
	\caption{Normalized effective throughput comparisons where we considered the EP-based scheme with QPSK, 8-PSK, and 16-QAM.}
        \label{fig:T}
\end{figure*}
Thirdly, Fig.~\ref{fig:T} compared the normalized effective throughput given by \eqref{throughput}, where we considered the EP-based scheme with QPSK, 8-PSK, and 16-QAM.
The effective throughput performance under the PCSI assumption was also plotted as a reference.
In Fig.~\ref{fig:T_coderate}, we considered the equal coding rate $r_c=0.75$, and the different transmission rate between the EP-based and SP-based schemes.
The number of data symbols $N_d$ for each scheme is given by
\begin{align}
\begin{cases}
N_{d,\,\mathrm{EP}} &= NM - (2\lmax + 1) (4\kmax + 1), \\
N_{d,\,\mathrm{SP}} &= NM.
\end{cases}
\label{transmission_rate}
\end{align}
Then, the SP-based schemes improved the transmission rate by $56\%$ in our simulation parameters.
As a result, except for the result assuming PCSI, the EP-based scheme with QPSK showed the throughput gain in the low SNR region, while the multiple SP-based scheme with the coded iteration process achieved the best performance in the intermediate SNR region.

Given a specific target rate for each scheme, which one achieves the largest gain?
To investigate the best scheme in terms of throughput, Fig.~\ref{fig:T_transmissionrate} compared the normalized effective throughput, where we adjusted the coding rates among the EP-based and SP-based schemes to match the transmission rate to almost $1.5~\mathrm{bit/sym}$.
Through the transmission rate matching, the coding rates of the EP-based scheme with 8-PSK and 16-QAM were determined as $r_c \approx 0.78$ and $r_c \approx 0.59$, respectively.
The EP-based scheme with QPSK was also evaluated in Fig.~\ref{fig:T_transmissionrate} as a benchmark, but it exceptionally assumed the uncoded scenario because the transmission rate in that case was less than $1.5~ \mathrm{bit/sym}$ despite $r_c=1$.

Under the equal transmission rate assumption, the multiple SP-based scheme with $N_p=9$ achieved the best throughput performance in entire SNR region.
In terms of the effective throughput, it can be seen from Fig.~\ref{fig:T_transmissionrate} that maximizing the number of pilots $N_p$ for the multiple SP-based scheme is the best solution, compared to adjust the modulation order and coding rate for the EP-based scheme.
We also found that the effective throughput is further enhanced by employing the proposed algorithm under the optimal parameter setting demonstrated in Fig.~\ref{fig:NMSE_runc}.

\subsection{Effects of the Increased Modulation Order \label{subsec:16-qam}}
\begin{figure}[tbp]
    \centering
    \includegraphics[clip, scale=0.64]{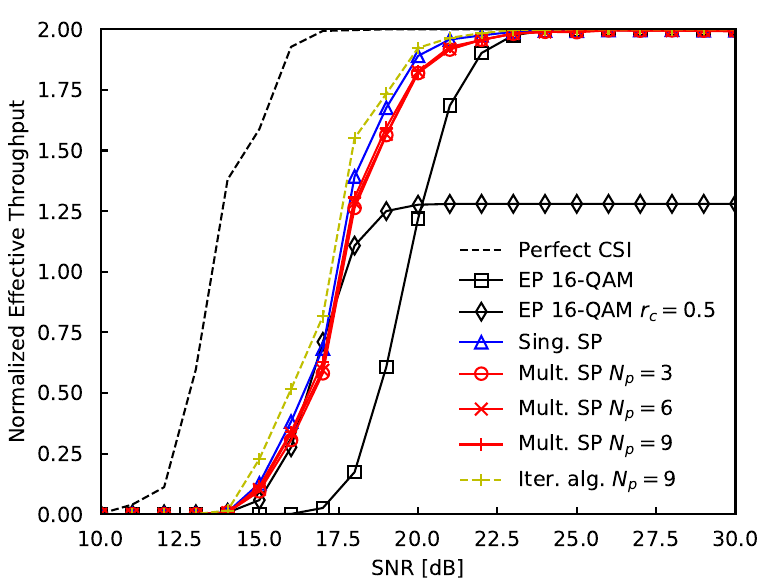}
    \caption{Normalized effective throughput comparison with 16-QAM signaling.}
    \label{fig:T_transmissionrate_16-qam}
\end{figure}
Through a performance comparison with 16-QAM signaling $(L = 16)$, we investigated the effects of the increased modulation order. To compensate for the reduced reliability caused by the higher modulation order, we set the coding rate $r_c = 0.5$ for the SP-based schemes. As for the EP-based scheme, we set $r_c = 0.5$ as well, and in order to match the transmission rates among all schemes, we also set $r_c \approx 0.78$.

Finally, Fig.~\ref{fig:T_transmissionrate_16-qam} compared the normalized effective throughput with 16-QAM signaling.
To match the transmission rates among all schemes, we set the coding rate $r_c \approx 0.78$ for the EP-based scheme.
As shown in Fig.~\ref{fig:T_transmissionrate_16-qam}, the SP-based schemes achieved better throughput compared to the EP-based scheme, even in the case of 16-QAM signaling.
Focusing on the SP-based schemes with the conventional algorithm, the case of single SP showed the best throughput, which indicates that the multiple SP-based schemes reduce interference cancellation performance due to the increased modulation order.
Contrary, the SP-based scheme achieved throughput improvement by employing the iterative algorithm with symbol replica generation.
This result demonstrated that the advantage of the proposed algorithm is independent of the modulation order.

\section{Conclusions \label{sec:conc}}
In this paper, we proposed the novel pilot assignment scheme with multiple SPs and iterative algorithm suitable for coded OTFS systems.
While the conventional single SP-based scheme suffered serious data-pilot interference in the high SNR region, the proposed scheme with multiple SPs realized the channel estimation and interference cancellation simultaneously, resulting in the improved throughput.
Contrary to the EP-based scheme, the multiple SP-based scheme did not require the guard space, which also contributed to the increased throughput.
Focusing on the property that the quality of the interference cancellation relies on the accuracy of data symbols, we incorporated the channel decoder into the iterative algorithm.
As a result, utilizing the refined symbol replicas based on the LLR information, the proposed algorithm further enhanced the throughput performance.
In addition, based on the unified transmission rate and fair SNR definition, we demonstrated the advantage of the proposed pilot assignment scheme and iterative algorithm in terms of the effective throughput, compared with the conventional EP-based and single SP-based schemes, with QPSK and 16-QAM signaling.

A key advantage of the proposed system is its support for a tunable complexity-performance trade-off via the iteration parameters $\{r_{\text{unc.}},, r_{\text{cod.}},, r_{\text{end}}\}$.
This flexibility allows the receiver to balance complexity and throughput.
As a promising direction for future work, an adaptive stopping criterion can be developed to select these parameters dynamically.
For example, early termination based on LLR variance or residual interference power could enable the receiver to allocate computational effort only when justified by the channel conditions or SNR level.

Another open problem is threshold design for the EP- and SP-based estimators in fractional-Doppler scenarios. Non-integer Doppler shifts spread energy across delay-Doppler bins, and a fixed threshold can fail in some cases. Although we have verified that reducing the modulation order alleviates this issue thanks to the iterative decoding algorithm, it is common to all current schemes, and another strategy remains an important target for future research.

\section*{Acknowledgment}
The authors would like to thank Dr. Yuto Hama from Ericsson Research, Japan, for providing valuable comments.

\footnotesize{
    \bibliographystyle{IEEEtranURLandMonthDiactivated}
	\bibliography{main}
}

\begin{IEEEbiography}[{\includegraphics[width=1in,height=1.25in,clip,keepaspectratio]{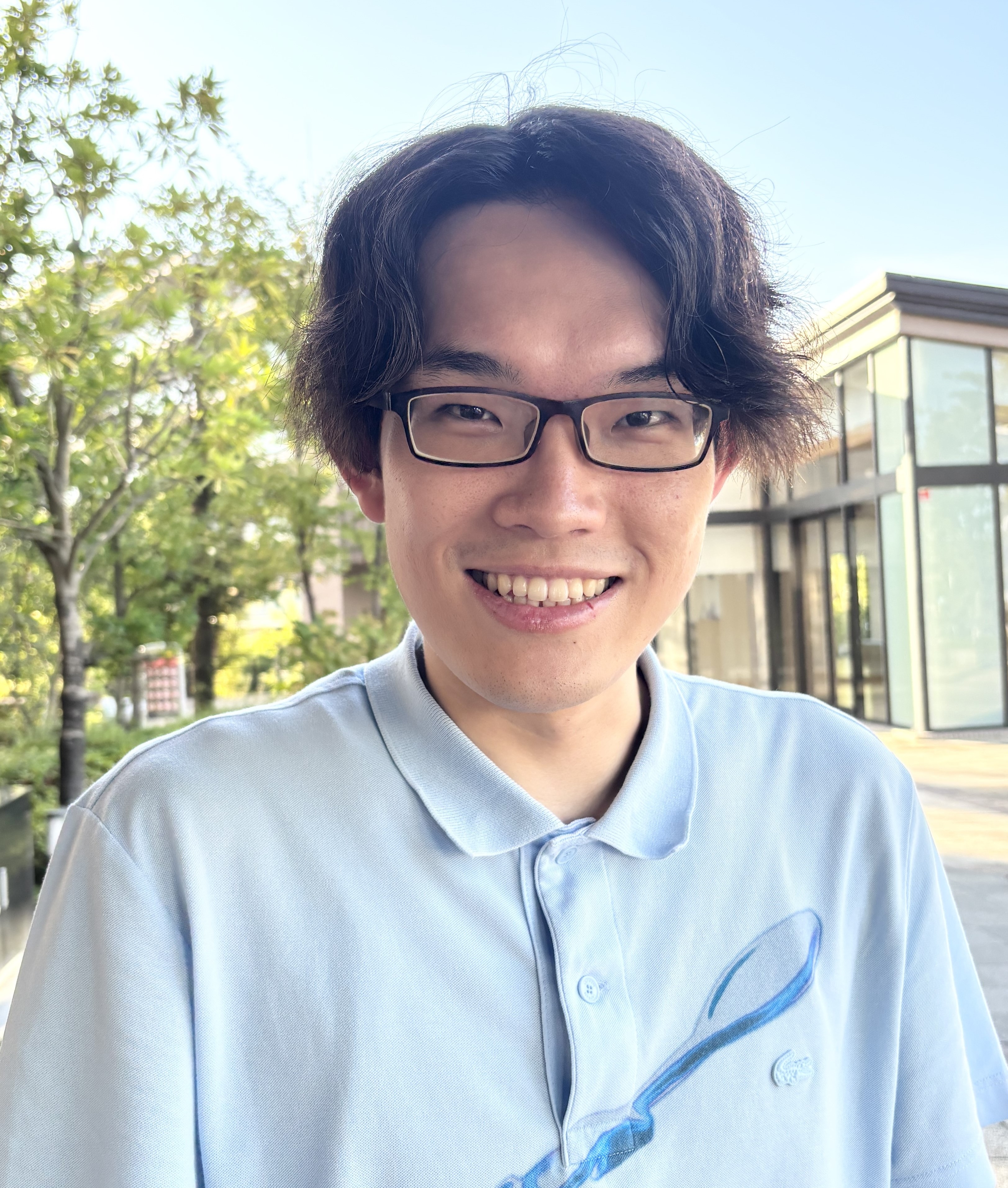}}]{Yuta Kanazawa} 
received the B.E. and M.E. degrees from Yokohama National University, Kanagawa, Japan, in 2023 and 2025, respectively. He received the IEEE VTS Tokyo/Japan Chapter Young Researcher's Encouragement Award in 2023. He is currently working at Ericsson Japan. His research interests include OTFS and channel estimation.
\end{IEEEbiography}

\begin{IEEEbiography}[{\includegraphics[width=1in,height=1.25in,clip,keepaspectratio]{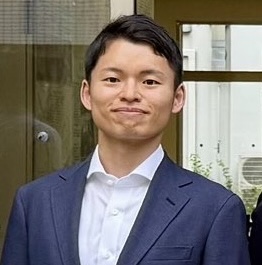}}]{Hiroki Iimori} (S'18--M'22) is a Senior Researcher at Ericsson Research. He earned his Ph.D. with special distinction (summa cum laude) from Jacobs University Bremen, Germany, in 2022, during which he worked as a Research Associate on a collaborative project with Continental AG (2019–2022). He was also a Guest Researcher at the University of Electro-Communications, Japan, and a Visiting Student at the University of Toronto, Canada, in 2020. He holds B.Eng. and M.Eng. (Hons.) degrees in Electrical and Electronic Engineering from Ritsumeikan University, Kyoto, Japan, awarded in 2017 and 2019. His research interests include optimization theory, wireless communications, and signal processing. His honors include the Ericsson Key Contributor Award (2025), the IEICE Young Researcher of the Year Award (2020), and the Yoshida Scholarship Foundation Doctoral Fellowship (2019–2022), among others.
\end{IEEEbiography}

\begin{IEEEbiography}[{\includegraphics[width=1in,height=1.25in,clip,keepaspectratio]{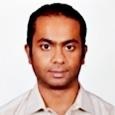}}]{Chandan Pradhan} received the B.Tech. degree from IIIT Bhubaneswar, India, in 2013, and the M.S. degree from IIIT Hyderabad, India, in 2016, and the Ph.D. degree form The University of Sydney, Sydney, NSW, Australia, in 2021. He has worked as Research Engineer in CEWiT, India, in 2016. He is currently working as Senior Researcher in Ericsson research, Japan. His research interests include beamforming techniques in MIMO systems and application of deep-learning in wireless networks.
\end{IEEEbiography}

\begin{IEEEbiography}[{\includegraphics[width=1in,height=1.25in,clip,keepaspectratio]{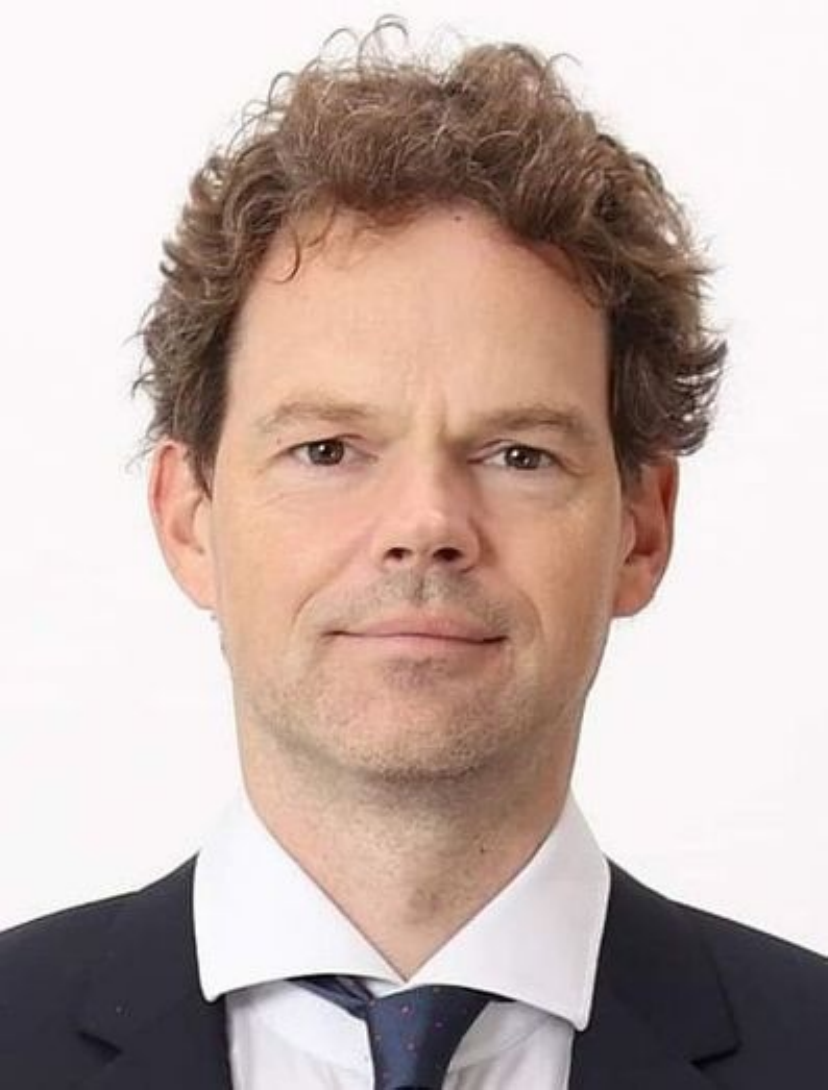}}]{Szabolcs Malomsoky} is the head of Ericsson Research Japan. Earlier he was leading research units in Hungary, Sweden and the US. He received a Ph.D. degree from the Budapest University of Technology and Economics in 2003. Szabolcs worked with strategy setting and technical leadership in research areas including wireless technologies, network analytics, cloud computing, network management and programmable networks.

\end{IEEEbiography}

\begin{IEEEbiography}[{\includegraphics[width=1in,height=1.5in,clip,keepaspectratio]{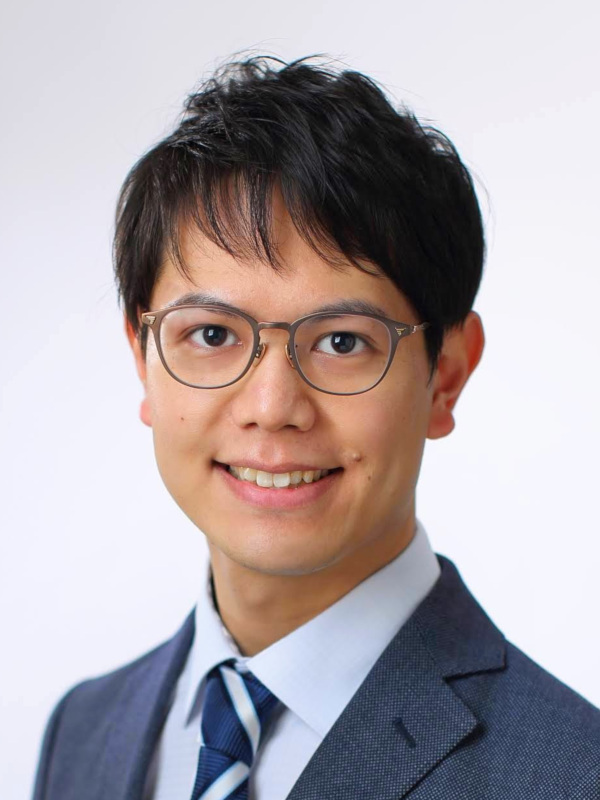}}]{Naoki~Ishikawa}
(S'13--M'17--SM'22) is an Associate Professor with the Faculty of Engineering, Yokohama National University, Kanagawa, Japan. He received the B.E., M.E., and Ph.D. degrees from the Tokyo University of Agriculture and Technology, Tokyo, Japan, in 2014, 2015, and 2017, respectively. In 2015, he was an Academic Visitor with the School of Electronics and Computer Science, University of Southampton, UK. From 2016 to 2017, he was a research fellow of the Japan Society for the Promotion of Science. From 2017 to 2020, he was an assistant professor in the Graduate School of Information Sciences, Hiroshima City University, Japan. Since 2025, he has been a Research Scholar with the Electrical and Computer Engineering, University of California, San Diego, US. He was certified as an Exemplary Reviewer of \textsc{IEEE Transactions on Communications} in 2017 and 2021. He has served as Associate Editor of \textsc{IEEE Transactions on Vehicular Technology} since 2025. His research interests include quantum algorithms and wireless communications.
\end{IEEEbiography}

\end{document}